# A Blockchain-Based Approach for Secure and Transparent e-Faktur Issuance in Indonesia's VAT Reporting System

Ghardy Lazuardy Farchan
*Information Systems Management Department, BINUS Graduate Program - Master of Information Systems Management*
Bina Nusantara University
Jakarta, Indonesia 11480
ghardy.farchan@binus.ac.id

*Abstract*— The implementation of blockchain technology in tax administration offers promising improvements in security, transparency, and efficiency. This paper presents the design of a blockchain-based e-Faktur system aimed at addressing the challenges of issuing and verifying tax invoices within Indonesia's VAT reporting process. Utilizing Hyperledger Fabric, a private permissioned blockchain, integrated with Hyperledger FireFly, this system ensures data immutability, access control, and decentralized transaction processing. The proposed system streamlines the issuance of NSFP and validates invoices while reducing reliance on centralized servers, eliminating single points of failure, and enhancing auditability. Hypothetical cyberattack scenarios were explored to assess the system's robustness. The results demonstrate that blockchain technology mitigates potential vulnerabilities and improves the resilience of tax reporting systems. The findings highlight the feasibility of blockchain in public tax administration and provide a foundation for future research on large-scale implementations.

*Keywords*— Blockchain, Hyperledger Fabric, e-Faktur, VAT reporting, tax administration, cybersecurity.

I. INTRODUCTION

Indonesia's VAT system faces significant challenges, with tax fraud and inefficiencies leading to substantial revenue loss. In 2021, the VAT c-efficiency rate was approximately 63.58%, meaning the country could only collect 63.58% of its potential VAT revenue [1]. This inefficiency is partially attributed to the presence of fraudulent tax invoices, which are often used by entities not verified as taxable enterprises or for fictitious transactions aimed at reducing tax liabilities or obtaining illegal refunds [2].

According to the 2021 Annual Report from the Directorate General of Taxes [3], tax-related crimes resulted in a loss of IDR 1.34 trillion in national revenue, with 43 of the 103 cases involving the issuance or use of fraudulent tax invoices. These statistics underscore the urgency of addressing fraud and inefficiencies within the VAT system to improve revenue collection and ensure tax compliance.

In response to these challenges, the Directorate General of Taxes introduced modern systems such as e-Nofa (electronic invoice numbering) and e-Faktur (electronic invoicing), which have been effective in reducing the number of fictitious tax invoice cases [4]. However, these systems remain centralized, making them vulnerable to manipulation and cyber-attacks.

To mitigate these weaknesses, an enhanced tax administration system is required to achieve regulatory compliance and maximize VAT potential [5]. Blockchain technology offers a promising solution to these vulnerabilities. Its decentralized ledger, which stores data across multiple nodes, significantly reduces the risk of manipulation or system failure. By ensuring that transactions are validated across various nodes, blockchain enhances data integrity and transparency [6], [7].

Blockchain technology can increase data security and reduce manipulation risks, even under high system demand [8]. By facilitating real-time transactions through smart contracts, blockchain reduces administrative burdens, expedites payments, and improves auditing and fraud detection capabilities. The decentralized nature of blockchain prevents tax manipulation, corruption, and data forgery, while enhancing the efficiency and transparency of tax reporting [9], with the potential to revolutionize tax administration [10].

Despite its promise, research on blockchain's application in Indonesia's VAT system remains limited. While many theoretical frameworks have been proposed, few studies have progressed to the prototype or implementation stage [11]. This work aims to address this gap by presenting a blockchain-based approach focused on the issuance process of e-Faktur. Ensuring that tax invoices are legitimate and tamper-proof is essential for enhancing security and transparency in Indonesia's VAT reporting system.

The primary objective of this research is to explore how blockchain technology can enhance the security, transparency, and integrity of Indonesia's VAT system, particularly in the issuance of tax invoices. This paper also presents a system prototype and evaluates its robustness through hypothetical scenarios, assessing its effectiveness in preventing fraudulent activities in tax administration.

## II. RELATED WORK

Blockchain technology has been increasingly explored as a means to improve tax systems, particularly in ensuring transparency, security, and efficiency in VAT administration. Several studies have focused on its potential to enhance tax oversight while addressing systemic inefficiencies.

Reference [12] examined the potential of blockchain in VAT systems, particularly in Indonesia. That study highlights how blockchain could be implemented in the Tax Invoice Serial Number system, enabling real-time tracking and monitoring of transactions by the Directorate General of Taxes. The use of a permissioned private blockchain enhances both security and transparency in the system by allowing greater control over access. In contrast, the system proposed in this work integrates both issuance and validation of VAT invoices, offering a more comprehensive solution for fraud prevention within Indonesia's tax framework.

Reference [13] and [14] explored the theoretical benefits of blockchain for improving Indonesian tax administration, particularly in the context of combating fraud and enhancing administrative efficiency. Both studies discussed the potential for blockchain to decentralize tax reporting, reduce unauthorized access, and provide real-time access to transaction data for tax authorities. While these studies highlighted the advantages of blockchain in improving oversight and reducing administrative costs, they did not propose specific models or technical frameworks for practical implementation.

Internationally, reference [15] presented a blockchain-based VAT system for Saudi Arabia using Hyperledger Fabric, automating tax collection via smart contracts. While the Saudi system focuses on supply chain management, the proposed e-Faktur system in this work also utilizes Hyperledger Fabric but is specifically tailored to Indonesia's VAT system, targeting the prevention of fraudulent invoices. Although both systems use Hyperledger Fabric, their scope and application differ: the Saudi system is designed for large enterprises and supply chains, whereas the proposed system addresses the distinct challenges of transparency and fraud prevention in Indonesia's VAT environment, particularly for small and medium-sized businesses.

## III. PROPOSED SYSTEM

### A. Blockchain System Overview

In the blockchain-based e-Faktur system, two key participants play crucial roles: Pengusaha Kena Pajak (PKP), or taxable enterprises, and the Direktorat Jenderal Pajak (DJP), or Directorate General of Taxes.

- PKP is responsible for issuing tax invoices and submitting transaction data to the system. PKP plays a critical role in generating accurate and tamper-proof records of VAT transactions.

- DJP oversees the validation of these tax invoices and ensures that the data submitted by PKP complies with Indonesia's tax regulations. Through blockchain's real-time auditing capabilities, DJP can monitor the integrity of the system.

By design, blockchain ensures data immutability—once a transaction is recorded, it cannot be altered without detection. This feature secures tax invoice data and enhances the trustworthiness of the system.

This interaction between PKP and DJP also ensures decentralization, as tax invoice records are verified across multiple entities without relying on a centralized authority. While PKP is responsible for creating tax invoices, DJP's role in validating them prevents any single entity from having full control over the system.

To determine the most appropriate blockchain solution for the e-Faktur system, we applied the blockchain decision tree presented in [16]. The decision-making process involved the following considerations:

1. Is there a need to store transaction state? Yes, the system must store transaction states for audit purposes and historical record-keeping, ensuring that tax invoice data is accurately tracked over time.

2. Are there multiple writers? Yes, both PKP and DJP are responsible for issuing and verifying tax invoices, making it necessary to support multiple participants who write data to the blockchain.

3. Can an always-online trusted third party (TTP) be used? No, relying on a trusted third party introduces risks of centralization and single points of failure. Blockchain removes this risk by decentralizing control, ensuring that no single entity can alter data without the consensus of other participants.

4. Are all writers known? Yes, all participants—PKP and DJP—are known and identifiable. Identity verification is managed through digital certificates issued by a certificate authority, ensuring that only authorized entities can participate.

5. Are all writers trusted? No, while participants are known, they cannot all be fully trusted to perform their roles without oversight. The permissioned blockchain uses cryptographic signatures and certificates to ensure that transactions are verified and trust is maintained between participants, without requiring full trust in any single entity.

6. Is public verifiability required? No, tax-related data is confidential, so public verifiability is unnecessary. Access is restricted to authorized participants only.

Based on this analysis, a private permissioned blockchain was selected as the optimal solution. This approach ensures that only authorized entities can participate in the system, safeguarding sensitive tax information while maintaining data integrity.

A private permissioned blockchain offers superior control and security by restricting access to authorized entities, such as PKP and DJP in the e-Faktur system. The system enforces strict access control through role-based permissions and smart contracts. Each participant has specific permissions: PKP submits tax invoices, while DJP verifies and validates these transactions, ensuring that only authorized parties can interact with the system. This maintains the integrity of the tax records.

## B. Hyperledger Fabric and FireFly

Hyperledger Fabric was chosen for its flexibility and robust architecture, which is specifically designed for enterprise-level applications like Indonesia's VAT system. Its modular architecture allows components like the Ordering Service, Membership Service Provider (MSP), and Smart Contracts (referred to as Chaincode in Hyperledger Fabric) to be customized to meet the specific needs of VAT administration. This flexibility ensures that only authorized participants, such as PKP and DJP, can interact with the system securely, while maintaining data integrity and privacy through permissioned access.

A key benefit of using Hyperledger Fabric is its integration with Hyperledger FireFly, a framework designed to simplify blockchain application development by automating tasks such as smart contract creation and data management. This allows developers to focus on solving business challenges—such as ensuring efficient tax invoice issuance and management—without dealing with the complexities of blockchain infrastructure.

FireFly also enhances system performance by handling off-chain data, managing API integrations, and providing real-time monitoring capabilities. These features are critical for managing the high transaction volumes of the e-Faktur system, ensuring that the system remains scalable and responsive under load while maintaining secure communication between participants. Additionally, FireFly supports multi-party workflows and simplifies the coordination between on-chain and off-chain processes, which are essential for the proper functioning of e-Faktur.

Furthermore, the Blockchain-as-a-Service (BaaS) platform Kaleido was chosen during the prototyping phase due to its free access for development and testing. Kaleido supports the Raft consensus protocol, which is lightweight and crash fault-tolerant, making it well-suited for a permissioned system where participants (PKP and DJP) are known and verified. This ensures fast transaction processing while maintaining data integrity. While Kaleido was selected for its cost-effectiveness in prototyping, the system is flexible enough to be deployed on on-premise or cloud infrastructures in real-world implementations.

Together, Hyperledger Fabric and FireFly provide a scalable and secure solution for the issuance of e-Faktur, ensuring that Indonesia's VAT system can handle large transaction volumes, maintain strict access control, and deliver auditable records. This combination enhances security, transparency, and efficiency, making it an ideal fit for Indonesia's tax administration system.

## C. Business Process

The e-Faktur blockchain-based system focuses on automating and streamlining the issuance of tax invoices (e-Faktur) and the verification of Tax Invoice Serial Number (Nomor Seri Faktur Pajak or NSFP) by the Directorate General of Taxes. The following steps describe the new business process in the proposed system:

1. NSFP Request by PKP: The PKP submits an NSFP request via the e-Faktur system, providing necessary information such as the tax year and the required number of NSFPs. Once the request is submitted, the system automatically processes it and moves to the verification step.

2. NSFP Verification and Issuance via Chaincode: The chaincode processes the NSFP request by verifying the PKP's eligibility. Once the NSFP is issued, the system notifies the PKP through the interface, automatically directing them to the next step where they can select the NSFP for invoicing. The issuance is recorded immutably on the blockchain, while the full NSFP data is securely stored in the system.

3. NSFP Selection for e-Faktur Issuance: PKP selects the issued NSFP to access the e-Faktur input section in the system. The selection of the NSFP automatically triggers the input of invoice data for tax invoicing.

4. Input of Invoice Data: PKP inputs essential transaction details such as transaction date, item details, prices, and the VAT amount. The system ensures accurate and consistent data input through predefined fields and formats. Once the data is submitted, the system automatically validates the input and moves to the validation and storage step.

5. Validation and Transaction Storage via Chaincode: The system automatically validates the invoice data and stores a hash of the invoice on the blockchain to maintain integrity, while securely storing the full invoice details within the system. In future implementations, this validation process will be expanded to perform checks such as verifying the NSFP ownership and ensuring compliance with tax regulations.

6. Print e-Faktur Document: Upon successful validation, PKP can print the e-Faktur document. The system retrieves the invoice data from secure storage and includes a blockchain verification hash in the printed document. This ensures the authenticity of the e-Faktur, while keeping the sensitive data secure and accessible only to authorized participants.

This integration of the NSFP request and e-Faktur issuance into a unified workflow ensures the efficiency, security, and transparency of the entire tax invoicing process.

## D. User Interaction with the System

The blockchain-based e-Faktur system involves two main actors: PKP and DJP. Their interactions with the system are described below:

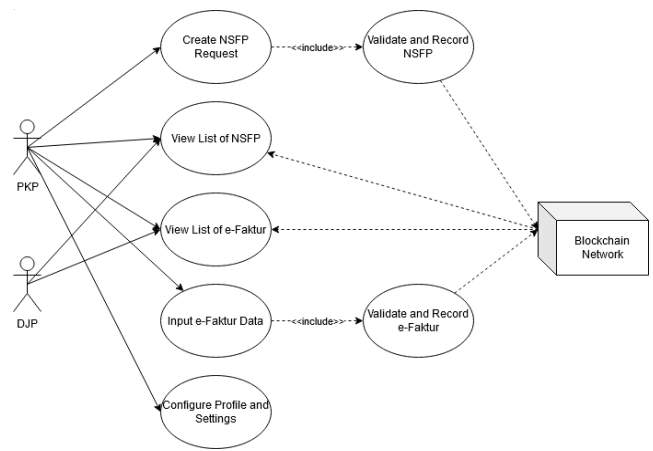

Fig. 1. e-Faktur System Use Case Diagram

This use case diagram illustrates their interactions with the system, such as requesting and validating NSFP, inputting invoice data, viewing NSFP and e-Faktur, and configuring profile and system settings. Blockchain technology plays a central role in validating and recording these transactions immutably using chaincode. This ensures data integrity, as once data is recorded on the blockchain, it cannot be altered.

The system employs Raft consensus protocol to validate transactions, ensuring that all participants agree on the data added to the blockchain. Furthermore, each transaction is timestamped and auditable, providing transparency and accountability for both PKP and DJP.

The "Configure Profile and Settings" use case allows PKP to update their user profile information, such as address or tax identification number, and adjust system preferences including connecting to the FireFly API.

## IV. IMPLEMENTATION

### A. Participants and Data Handling

In the e-Faktur blockchain system, both PKP and DJP act as participants, each identified through a certificate authority (CA). This ensures that only authorized entities can interact with the system. The system manages two primary types of data assets: NSFP and invoices (Faktur). Data handling within the system follows two key mechanisms for data exchange:

1. Broadcast Data: Public data, such as NSFP (Nomor Seri Faktur Pajak), is broadcast across the entire blockchain network, ensuring transparency. This allows all participants, including DJP and all PKP entities, to access the NSFP data. Broadcasting NSFP enables PKP to use the data for invoice creation, while DJP can monitor the issuance of NSFP in real-time.

2. Private Data Exchange: Sensitive information, such as invoice details in the Faktur, is exchanged privately between the submitting PKP and DJP. The system uses a private data exchange to ensure confidentiality, where only the originating PKP and DJP have access to these transactions. This safeguards the privacy of tax-related information, restricting access to only the necessary parties.

In the e-Faktur system, NSFP data is less sensitive and is therefore handled through broadcast exchanges, while Faktur data, containing transaction details like goods, prices, and VAT amounts, is managed through private exchanges to maintain the security and privacy of financial records.

PKP is responsible for both creating and submitting NSFP requests and Faktur data, with full read and write access to these assets. DJP oversees the validation and verification process, maintaining full read access to both NSFP and Faktur data. While DJP does not modify the data, its primary role is to ensure compliance and conduct audits of the records.

### B. Transactions

The blockchain-based e-Faktur system enables several key transactions that are processed via chaincode within FireFly API. These transactions allow participants to interact with both public and private data in the system.

TABLE I. LIST OF FIREFLY API

| Transaction | Participant | Description | Data Handling |
|---|---|---|---|
| GetNsfp | PKP, DJP | Retrieve all issued NSFPs. | Broadcast |
| GetFaktur | PKP, DJP | Retrieve issued Faktur invoices. | Private Data Exchange |
| PostNsfp | PKP | Request and issue NSFPs. | Broadcast |
| PostFaktur | PKP | Submit invoice data for validation. | Private Data Exchange |

These transactions provide the backbone of the e-Faktur system, ensuring that NSFP data is openly shared to enable transparency, while Faktur data is protected through secure, private exchanges.

### C. Data Storage

While FireFly facilitates smart contract execution and data exchanges, the system separates on-chain and off-chain data storage for efficiency. On-chain transactions record only hashes of the data, ensuring immutability and traceability. The actual data is stored off-chain due to the potentially large size of invoices and the inefficiency of storing this data directly on the blockchain.

- NSFP data is stored on the InterPlanetary File System (IPFS), which provides decentralized storage. Only the hash of the data is stored on-chain for verification, ensuring data integrity. While NSFP data is considered less confidential, IPFS ensures decentralized distribution, making it retrievable by authorized participants within the e-Faktur system.

- Faktur data is securely exchanged between the PKP (sender) and DJP (receiver). Transmission is protected by end-to-end encryption and Mutual TLS authentication, ensuring secure and verified communication. Both parties store the private data in their respective databases, while a cryptographic hash of the data is recorded on-chain to maintain immutability.

This separation of data storage ensures that sensitive information remains secure while maintaining the benefits of blockchain's immutability and transparency for critical public data like NSFP. FireFly synchronizes on-chain and off-chain data using a global event sequencing system, where public and private payloads are processed separately but linked through hashes. Only the originating PKP and DJP can access private data, as the synchronization ensures selective visibility between participants.

### D. System Architecture

The e-Faktur system leverages Hyperledger FireFly for scalability and efficient management of enterprise-grade workloads. Built on Kubernetes, FireFly enables horizontal scaling, allowing the system to handle increasing transaction volumes by distributing tasks across multiple nodes. FireFly's multi-chain and multi-protocol support also facilitates seamless integration with Hyperledger Fabric, which underpins the decentralized structure of the e-Faktur system. This support allows the system to handle complex, multi-party interactions between DJP and various PKP nodes.

The system operates in a decentralized manner, with each participant (e.g., DJP and various PKP nodes) running its own components, such as the front-end application, FireFly, orderer node, peer node, certificate authority (CA), and IPFS.

FireFly's messaging mechanisms ensure secure and efficient communication between these nodes. FireFly's architecture supports fault tolerance through cloud deployment with redundancy and backup mechanisms, ensuring continuous operation even in the event of system failures.

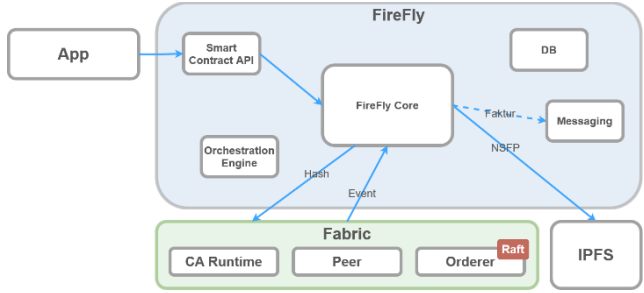

Fig. 2. e-Faktur System Node Diagram

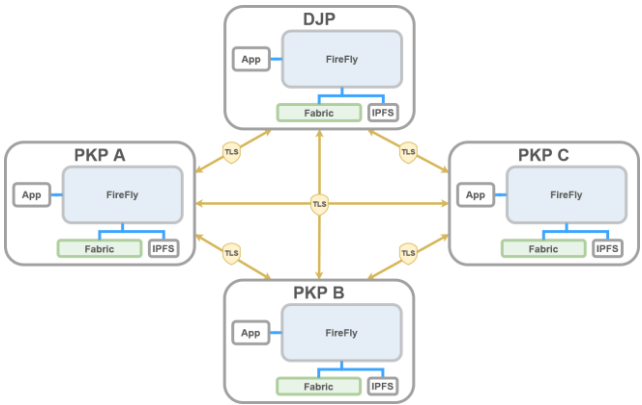

Fig. 3. e-Faktur System Network Diagram

*E. Hypothetical Cyberattack Scenarios and System Resilience*

To explore the robustness and potential response of the blockchain-based e-Faktur system, several hypothetical cyberattack scenarios were devised. These simulations are purely theoretical and serve as an initial exploration of the system's response to common cyberattacks. The following types of attacks were considered:

TABLE II. TYPE OF CYBERATTACKS

| Attack Type | Description |
| --- | --- |
| Phishing | Attackers attempt to steal credentials from PKP or DJP employees through malicious emails containing harmful links. |
| SQL Injection | Exploiting input form vulnerabilities to insert malicious SQL commands, aiming to manipulate or create fake invoices. |
| Man-in-the-Middle (MITM) | Attackers intercept and modify invoice data during transmission to alter transaction details. |
| Ransomware | Attackers encrypt data within the e-Faktur system, demanding a ransom for decryption, potentially affecting data integrity. |

These theoretical simulations assess the blockchain system's potential to preserve data integrity and ensure operational continuity during attacks. However, practical testing has not been conducted at this stage, and real-world simulations will be necessary to further validate the system's effectiveness. Future research should focus on performing these tests under real-world conditions to ensure the system's robustness and scalability.

In these simulated scenarios, the blockchain-based e-Faktur system demonstrates strong resilience due to its decentralized and immutable ledger. During a hypothetical cyberattack, the on-chain data—comprised of cryptographic hashes of invoices—remains secure and unchangeable. Hyperledger Fabric and FireFly would assist in detecting suspicious activity using smart contracts, which monitor for anomalies like unauthorized transaction modifications.

Upon detection of suspicious activity, the system could isolate compromised nodes by revoking their digital certificates, preventing further network infiltration.

For off-chain data, recovery is handled through backups, with integrity verification ensured by comparing the recovered data to the original on-chain hashes. Sensitive invoice data is stored within private data collections, ensuring that it is protected and verifiable.

Throughout the process, all events—from attack detection to data recovery—would be recorded in the blockchain, offering complete transparency and auditability. Without blockchain, the system would be more vulnerable to manipulation and less efficient in data recovery. Blockchain technology preserves data integrity and reduces downtime.

While these hypothetical simulations suggest that the system would successfully mitigate cyberattacks, future work should include practical testing to validate its robustness under real-world conditions. This could involve performance benchmarking tools like Hyperledger Caliper to evaluate system scalability and resilience in a live environment, ensuring more concrete validation of the system's security and operational continuity.

In these theoretical scenarios, the blockchain-based e-Faktur system is expected to successfully mitigate all simulated cyberattacks. Data encrypted or manipulated by attackers could be recovered and verified through the blockchain, ensuring its authenticity.

Additionally, the entire response process—from detection to recovery—would be recorded on the blockchain, providing transparency. The blockchain-enabled system is designed to maintain operational continuity with minimal disruption, demonstrating the effectiveness of decentralization and secure data management.

## V. CONCLUSION

This research proposed and designed a blockchain-based e-Faktur system aimed at improving the issuance and verification of tax invoices within Indonesia's VAT reporting framework. Utilizing Hyperledger Fabric, a private permissioned blockchain, the system enhances security and transparency through an immutable ledger accessible only by authorized entities, such as the Directorate General of Taxes and taxable enterprises. This ensures data integrity and mitigates the risk of fraudulent tax invoices. Additionally, the decentralized architecture reduces the load on central servers, eliminates single points of failure, and increases system reliability.

While the focus of this study was on the issuance of output tax invoices, the flexibility of Hyperledger FireFly built on Hyperledger Fabric enables future scalability and modularity.

This ensures the system can adapt to large-scale implementation while maintaining privacy and security for sensitive data. However, the system's efficacy has yet to be tested under real-world conditions, and practical testing is necessary to fully evaluate its performance and security.

Future research should focus on large-scale testing to assess system performance under operational conditions. Expanding the blockchain framework to more complex taxation systems, such as income tax administration or luxury goods tax (PPnBM), could provide broader insights into the technology's applicability. Moreover, testing this system in other developing countries facing similar tax challenges may further demonstrate blockchain's ability to enhance tax administration on a global scale.